\documentclass[conference]{IEEEtran}
\IEEEoverridecommandlockouts
\usepackage{cite}
\usepackage{amsmath,amssymb,amsfonts}
\usepackage{algorithmic}
\usepackage{graphicx}
\usepackage{textcomp}
\usepackage{xcolor}
\usepackage{caption}
\usepackage{url}
\usepackage{subcaption}
\usepackage{fancyhdr}
\def\BibTeX{{\rm B\kern-.05em{\sc i\kern-.025em b}\kern-.08em
    T\kern-.1667em\lower.7ex\hbox{E}\kern-.125emX}}
\usepackage{fancyhdr}
\begin{document}
\bstctlcite{IEEEexample:BSTcontrol}

\newcommand{\TODO}[1]{\color{red}\textbf{[TODO: #1]}\color{black}}
\newcommand{\NOTICE}[1]{\color{blue}\textbf{[NOTICE: #1]}\color{black}}
\newcommand{\REFER}{\color{magenta}\textbf{[NEED REFERENCE]}\color{black}}
\newcommand{\etal}{{\it et al.}}

\captionsetup[subfigure]{labelformat=simple}
\renewcommand{\thesubfigure}{(\alph{subfigure})}

\title{Compound Virtual Screening by Learning-to-Rank\\ with Gradient Boosting Decision Tree and Enrichment-based Cumulative Gain}
\author{\IEEEauthorblockN{%
Kairi Furui}
\IEEEauthorblockA{\textit{Department of Computer Science} \\
\textit{School of Computing}\\
\textit{Tokyo Institute of Technology}\\
Kanagawa, Japan\\
furui@li.c.titech.ac.jp}
\and
\IEEEauthorblockN{%
Masahito Ohue}
\IEEEauthorblockA{\textit{Department of Computer Science} \\
\textit{School of Computing}\\
\textit{Tokyo Institute of Technology}\\
Kanagawa, Japan\\
ohue@c.titech.ac.jp}
}

\maketitle
\thispagestyle{plain}
 \fancypagestyle{plain}{
 \fancyhf{} 
  \fancyfoot[L]{  \vspace{-8mm}\small
978-1-6654-8462-6/22/\$31.00~\copyright2022~IEEE \\2022 IEEE Conference on Computational Intelligence in Bioinformatics and Computational Biology (CIBCB) ~DOI:10.1109/CIBCB55180.2022.9863032} 
 \renewcommand{\headrulewidth}{0pt}
 \renewcommand{\footrulewidth}{0pt}
 }

\begin{abstract}
Learning-to-rank, a machine learning technique widely used in information retrieval, has recently been applied to the problem of ligand-based virtual screening to accelerate the early stages of new drug development.
Ranking prediction models learn based on ordinal relationships, making them suitable for integrating assay data from various environments.
Existing studies of rank prediction in compound screening have generally used a learning-to-rank method called RankSVM.
However, they have not been compared with or validated against the gradient boosting decision tree (GBDT)-based learning-to-rank methods that have gained popularity recently.
Furthermore, although the ranking metric called Normalized Discounted Cumulative Gain (NDCG) is widely used in information retrieval, it only determines whether the predictions are better than those of other models. In other words, NDCG cannot recognize when a prediction model produces worse than random results.
Nevertheless, NDCG is still used in the performance evaluation of compound screening using learning-to-rank.
This study used the GBDT model with ranking loss functions, called lambdarank and lambdaloss, for ligand-based virtual screening; results were compared with existing RankSVM methods and GBDT models using regression.
We also proposed a new ranking metric, Normalized Enrichment Discounted Cumulative Gain (NEDCG), aiming to evaluate the goodness of ranking predictions properly.
In addition, the results showed that the GBDT model with learning-to-rank outperformed existing regression methods using GBDT and RankSVM on diverse datasets.
Finally, NEDCG showed that the predictions by regression were comparable to random predictions in multi-assay, multi-family datasets, demonstrating its usefulness for a more direct assessment of compound screening performance.

\end{abstract}


\section{Introduction}
The development cost and time required to obtain approval for a new drug increase every year, with some estimating a cost of \$2.6 billion per drug~\cite{mullard2014new}, and others reporting a development time of more than 10 years from identifying lead compounds to clinical trials~\cite{10.1093/bib/bbp023}.
Virtual screening (VS) is a process of computationally searching an extensive compound library for an active compound against a target protein in the early stage of new drug development.
Virtual screening technology helps in the discovery process of hit compounds~\cite{yan2020augmenting}.
For virtual screening in the drug repositioning context, the drug Edaravone (Radicava), FDA-approved for cardiovascular indications, was identified as a neurotrophic~\cite{eleuteri2017staged}.
Virtual screening with the FDA-approved drug dataset was also implemented for the Coronavirus Disease-2019 (COVID-19) virus, which has caused a pandemic since December 2019~\cite{kandeel2020virtual}.
Improving the predictive accuracy of virtual screening and extending its applicability to a wide range of activity data are essential to reducing the cost of developing new drugs.

Ligand-based virtual screening (LBVS) is a method that uses activity information already obtained from assays and mainly employs machine learning methods such as regression~\cite{ma2015deep,li2021structure} and classification~\cite{schneider2008gradual,nigsch2008ligand,adeshina2020machine} prediction.
Recently, a learning-to-rank method based on the ordinal relationship of activity values was proposed for virtual screening~\cite{wassermann2009searching,joachims2002optimizing,agarwal2010ranking,rathke2011structrank,zhang2015drug,suzuki2018pkrank,ohue2019spdrank,matsumoto2021ranking}.
This approach has two advantages, of which 
the first is that learning-to-rank is more accurate than regression for ordinal predictions~\cite{zhang2015drug}.
In drug discovery, potentially active compounds selected through virtual screening are biochemically assayed to determine whether they are active.
Therefore, the goal of virtual screening is not to predict the exact activity value but to list compounds that are even slightly more active at the top of the prediction.
For this reason, learning-to-rank, predicting based on order, is appropriate for virtual screening.
The second advantage is that learning-by-rank is dependent on ordering relationships in comparable groups, such as assay, making it easy to integrate experimental information from different situations.
Affinity indices based on biochemical assays, such as half the maximum inhibitory concentration ($\mathrm{IC}_{50}$), vary widely from one assay system to another. This aspect makes it challenging to integrate assay data from different environments using regression methods.
In learning-to-rank, the distribution of measurements in each assay need not be identical because the model learns based on the ordinal relationship among compounds within each assay~\cite{zhang2015drug}.

However, there are several problems with existing studies of virtual screening using learning-to-rank.
First, existing studies~\cite{agarwal2010ranking,zhang2015drug,suzuki2018pkrank,ohue2019spdrank,matsumoto2021ranking} mainly focus on RankSVM, a machine learning method used for ranking prediction, and do not evaluate the effectiveness of new methods.
Recently, a learning-to-rank method called LambdaMART has been broadly used in the information retrieval field~\cite{Burges}. 
LambdaMART is a machine learning method that learns a gradient boosting decision tree (GBDT) with a ranking loss function. This method has attracted attention in machine learning competitions.
No comparisons have been reported regarding whether LambdaMART is superior to RankSVM or whether learning-to-rank is superior to regression in order prediction, even in the GBDT model, for virtual screening.
Furthermore, previous studies~\cite{zhang2015drug,suzuki2018pkrank,ohue2019spdrank} have evaluated the performance of ranking prediction using a metric called Normalized Discounted Cumulative Gain (NDCG)~\cite{jarvelin2017ir,jarvelin2002cumulated,burges2005learning}, but the suitability of NDCG for LBVS has not been adequately discussed.
NDCG is a metric designed for use in learning-to-rank performance evaluation in information retrieval, which differs from the situation in virtual screening.
Specifically, NDCG reports a maximum value of 1 to predict that a sequence is perfectly correct. However, the value reported by NDCG does not express the extent to which a prediction represents an improvement.
In LBVS, enrichment, i.e., the improvement degree relative to the random prediction is essential.
However, existing studies use NDCG as a metric for evaluating virtual screening against ranking prediction without considering these differences.

In this study, we evaluate the VS performance of a new GBDT model with learning-to-rank and compare the performance with existing models. Moreover, we develop a metric that can evaluate the performance of ranking prediction appropriately.
The three main contributions of this study are:
\begin{itemize}
    \item We modified the conventional NDCG and proposed NEDCG (normalized enrichment DCG), a metric expressing the improvement ratio over random guess in ranking prediction.
    \item We applied for the first time the GBDT model with lambdarank loss function to the virtual screening problem.
    \item We validated NEDCG based on a dataset constructed for multiple situations where assay data was available. The results showed that the prediction accuracy of the proposed method using GBDT outperformed that of the conventional method based on learning-to-rank. In addition, the validation indicated that in some cases, the learning-to-rank was more effective. In others cases, the regression models were more accurate in prediction, depending on the dataset. 
\end{itemize}

\section{Materials and Methods}
\subsection{GBDT and LambdaMART}
GBDT is a machine learning algorithm that minimizes the cost function by iteratively ensembling weak prediction models using decision trees.
Moreover, GBDT is a widely used algorithm, and several effective implementations exist, including XGBoost~\cite{chen2016xgboost} and LightGBM~\cite{ke2017lightgbm}.
In this study, we used LightGBM as an implementation of GBDT.

LambdaMART~\cite{Burges} is a method for learning GBDT with a ranking loss function called lambdarank~\cite{burges2006learning},
designed to directly optimize the value of NDCG~\cite{jarvelin2017ir,jarvelin2002cumulated,burges2005learning}.
The NDCG for the top $K$ cases from a group of $N$ cases is as follows:
\begin{align}
    {\rm NDCG@}K &= \sum_{i=1}^K \frac{G_i}{D_i},\nonumber\\
    G_i &= \frac{gain_i}{{\rm maxDCG}}, \nonumber\\
    gain_i& = 2^{y_i}-1, \nonumber\\
    D_i &= \log_2(i+1), \nonumber
\end{align}
where $i$ is the predicted rank, $y_i$ is the label at $i$, and ${\rm maxDCG}$  is a normalization constant, which is the maximum discounted cumulative gain (${\rm DCG}$) when the ranking prediction is correct.
If the top of the ranking is correctly ordered, NDCG approaches its maximum value of 1, while at its worst, it has a minimum value of 0.
The loss of lambdarank for each group is as follows:
\begin{align}
    l_1(\boldsymbol{y}, \boldsymbol{s}) &= \sum_{y_i>y_j}\rho_{ij}\left|G_i-G_j\right|\log(1+e^{-\sigma(s_i-s_j)}),\nonumber\\
    \rho_{ij}&=\left|\frac{1}{D_i}-\frac{1}{D_j}\right|, \nonumber
\end{align}
where $s_i$ and $s_j$ are the predicted ranking scores. Lambdarank learns order relationships by penalizing based on $\Delta {\rm NDCG}_{ij}=\rho_{ij}|G_i-G_j|$, i.e., the difference in NDCG when rank $i$ is swapped with rank $j$.

Moreover, we experimented with a loss function based on the lambdaloss framework, called NDCGloss2~\cite{wang2018lambdaloss}, which minimizes a cost function called ${\rm NDCG_{cost}}$.
The lambdaloss function for each group and ${\rm NDCG_{cost}}$ is as follows:
\begin{align}
    l_2(\boldsymbol{y}, \boldsymbol{s}) &= \sum_{y_i>y_j}\delta_{ij}\left|G_i-G_j\right|\log(1+e^{-\sigma(s_i-s_j)}), \nonumber\\
    \delta_{ij}&=\left|\frac{1}{D_{|i-j|}}-\frac{1}{D_{|i-j|+1}} \right|,\nonumber\\
    {\rm NDCG_{cost}}&=\sum_{i=1}^N G_i - \sum_{i=1}^N \frac{G_i}{D_i}. \nonumber
\end{align}

\subsection{Normalized enrichment DCG}
NDCG is one of the primary metrics used to measure the performance of learning-to-rank models.
In particular, NDCG is appropriate for comparative analysis of different models and determining which hyperparameters are better. However, this method lacks information on the degree of improvement of the ranking predictions from the pre-training situation.
For example, an area under the receiver operating characteristic curve (AUROC) of 0.5, used in the classification task, implies that the model makes random predictions.
Therefore, we propose a new normalized enrichment DCG (NEDCG), inspired by AUROC, removing the effect of random prediction (pre-training state) from NDCG.
NEDCG for the top $K$ cases are defined as follows:
\begin{align}
    {\rm NEDCG@}K &= \frac{{\rm DCG@}K-{\rm randomDCG@}K}{{\rm maxDCG@}K-{\rm randomDCG@}K},\\
    {\rm DCG@}K &= \sum_{i=1}^K \frac{gain_i}{\log_2(i+1)},
\end{align}
where ${\rm randomDCG@}K$ is the discounted cumulative gain ${\rm DCG@}K$ for the top $K$ cases when predicting randomly.
When calculating ${{\rm randomDCG@}K}$, the average $gain_{mean}$ of the gains for a group of $N$ cases is used as follows:
\begin{align}
    {\rm randomDCG@}K &= \sum_{i=1}^K \frac{gain_{mean}}{\log_2(i+1)},\nonumber\\
    gain_{mean} &= \frac{1}{N}\sum_{j=1}^N gain_{j}.\nonumber
\end{align}

\section{Experiments}
\begin{figure}[tb]
    \centering
    \vspace{-12mm}
	\includegraphics[width=\columnwidth, bb= 0 0 940 600 ]{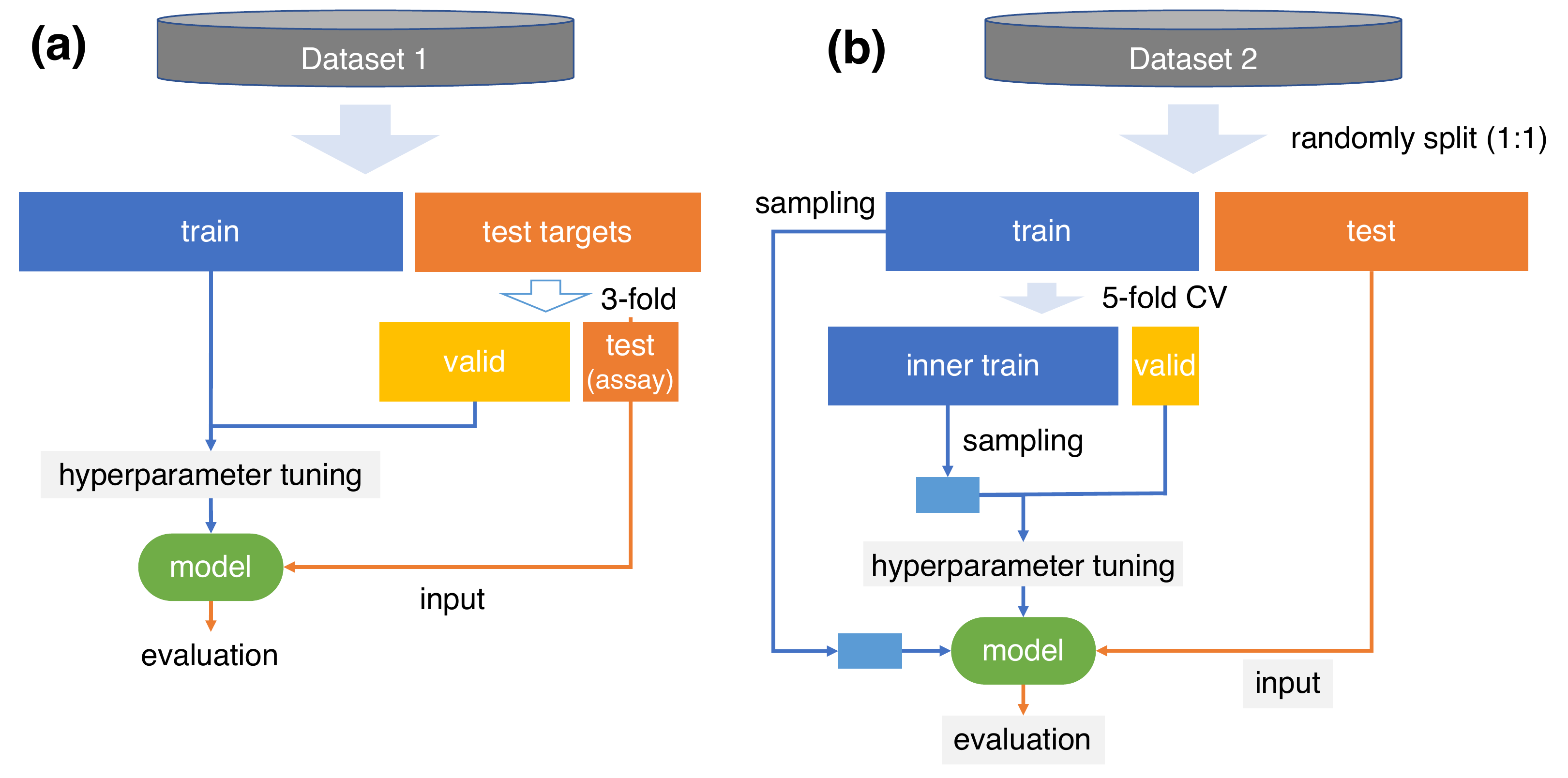}
    \caption{Experimental procedure for (a) Dataset 1 (complicated task) and (b) Dataset 2 (simple task).}
	\label{fig:concept}
\end{figure}
\subsection{Dataset}
In this section, we describe two datasets for different experimental situations.
Dataset 1 (complicated task) includes experimental data from various assays for different proteins of the same family are used as training data.
Dataset 2 (simple task) is a typical LBVS case where data from a single assay is randomly split into training and test data.

In Dataset 1, assay data for the phosphodiesterase (PDE) family were collected for $\mathrm{IC}_{50}$ from the ChEMBL database~\cite{10.1093/nar/gky1075} with reference to existing studies~\cite{zhang2015drug,suzuki2018pkrank}.
Data with the same assay ID were treated as a group, and assays with a group size of 5 or more were selected.
The objective variable was $\mathrm{IC}_{50}$ converted to $\mathrm{pIC}_{50}=-\log_{10}(\mathrm{IC}_{50})$.
For inactive data (cases with no value data were described as inactive), $\mathrm{pIC}_{50}=0$ based on~\cite{ohue2019spdrank}.
Table~\ref{tab:exp1_datasets} shows the details of Dataset 1.
In addition, Table \ref{tab:exp1_test} lists the ChEMBL Assay IDs of the assays used as test data and the number of compounds in each.

In Dataset 2, the Luciferase dataset (AID: 1006) for the inhibition rate measurement experiment was used following a previous study~\cite{ohue2019spdrank}.
The Luciferase dataset was obtained from PubChem BioAssay~\cite{wang2014pubchem}, a compound activity information database.
The activity value for Dataset 2 was $\%$Inhibition. However, because the gain used in the NDCG calculation was a power of 2 of the activity value, the inhibition rate value was divided by 10 and converted so that the maximum value is 10.
Note that the measured inhibition rate ranged from $-\infty$ to $100\%$, but the negative values were converted to set the inhibition rate to 0.
Dataset 2 contains 2,976 active compounds (inhibition rate $\geq 50$) and 192,588 inactive compounds (inhibition rate $< 50$).

\begin{table}[tbp]
    \centering
    \caption{Details of Dataset 1 (PDE family dataset). Target proteins in bold were excluded from the training step.}
    
    \label{tab:exp1_datasets}
    \begin{tabular}{c|r|r}
    \hline\hline
    protein name & \#{}assays & \#{}compounds \\ \hline
        PDE 1A & 4 & 76\\ 
        PDE 1B & 9 & 101\\ 
        PDE 1C & 7 & 106\\
        \textbf{PDE 2A} & 41 & 999\\
        PDE 3A & 22 & 324\\
        PDE 3B & 8 & 121\\
        PDE 4A & 36 & 685\\
        PDE 4B & 69 & 1,457\\
        PDE 4C & 9 & 110\\
        PDE 4D & 57 & 985\\
        PDE 5A & 87 & 2,889\\
        PDE 6A & 1 & 47\\
        PDE 6C & 3 & 33\\
        PDE 6D & 1 & 24\\
        \textbf{PDE 7A} & 22 & 659 \\
        PDE 7B & 2 & 15 \\
        PDE 8A & 4 & 43 \\
        \textbf{PDE 8B} & 2 & 168 \\
        PDE 9A & 23 & 568 \\
        PDE 10A & 79 & 3,848 \\
        PDE 11A & 12 & 181 \\ \hline
    \end{tabular}
\end{table}
\begin{table}[tbp]
    \centering
    \caption{Dataset 1: Assays used as tests and their sizes.}
    \label{tab:exp1_test}
    \begin{tabular}{l|l|r}
    \hline\hline
        ~ & ChEMBL Assay ID & \#{}compounds \\ \hline
        PDE 2A & CHEMBL3706318 & 146 \\ 
        PDE 7A & CHEMBL3706063 & 170 \\ 
        PDE 8B & CHEMBL2073616 & 111 \\ \hline
    \end{tabular}
\end{table}

\subsection{Procedure}
We compare the following 4 prediction models:
\begin{enumerate}
    \item lambdaloss (rank): GBDT ranking prediction model using the lambdaloss function
    \item lambdarank (rank): GBDT ranking prediction model using the lambdarank loss function
    \item RankSVM (rank): RBF Kernel RankSVM model based on PKRank~\cite{suzuki2018pkrank} for Dataset 1. Linear kernel RankSVM model based on SPDRank for Dataset 2.
    \item GBDT regression (regression): GBDT regression model with $L_2$ loss
\end{enumerate}

The experiments with regression were intended to compare the use of a ranking loss function with a regression loss function. The experiments with RankSVM were intended to compare the other learning-to-rank methods with the proposed method.

We adopted RankSVM as a representative of the existing learning-to-rank models because it was the best learning-to-rank model in a previous study~\cite{zhang2015drug}.
In Dataset 2, we used SPDRank, unlike Dataset 1, because training with the SPDRank model using stochastic gradient descent was successful.

\subsection{Features}
The 1-D and 2-D descriptors (1,613 dimensions) from mordred~\cite{moriwaki2018mordred} (Version: 1.2.0) were used as descriptors for the compounds.
However, we removed descriptors taking null values for more than half of the compounds in the training data, resulting in 1,452 dimensions in Dataset 1 and 1,447 dimensions in Dataset 2.
In addition, the normalized Smith-Waterman scores with targets in the training data were used as protein features.

However, for RankSVM in Dataset 2, we used ECFP4 (2,048 bit)~\cite{rogers2010extended} features computed with RDKit~\cite{rdkit} because SPDRank can learn only sparse features.

\subsection{Training and hyperparameters}
The number of rounds in the GBDT model was 100 when tuning the hyperparameters, and the number of rounds for validation was decided by \texttt{early\_stopping = 1000} based on NDCG@10.
The test data predictions were made using a model with 1.1 times the number of rounds determined for the validation data.
For tuning the GBDT hyperparameters, the number of leaves in the decision tree \texttt{num\_leaves} were (15, 31, 63, 127, 255, 511, 1023, 2047, 4095), and the minimum number of data assigned to a leaf \texttt{min\_data\_in\_leaf} were (10, 25, 50, 100, 200).
The other hyperparameters were set as follows: \texttt{lambda\_l1} = 0, \texttt{lambda\_l2} = 0, \texttt{feature\_fraction} = 0.7, \texttt{bagging\_fraction} = 1.0, and \texttt{bagging\_freq} = 0. 
For the learning rate, we used \texttt{learning\_rate} = 0.1 for hyperparameter tuning and \texttt{learning\_rate} = 0.05 for all other learning.

For Dataset 1, when using the lambdarank and lambdaloss loss functions in the GBDT model, the \texttt{lambdarank\_truncation\_\_level} was fixed at 30, and the \texttt{label\_gain} step width $\delta$ was fixed at 1.0.
For Dataset 2, when using the lambdarank and lambdaloss loss functions in the GBDT model, the \texttt{lambdarank\_truncation\_\_level} was fixed at 200, and the \texttt{label\_gain} step width $\delta$ was fixed at 0.01.

In the RankSVM experiments, the PKRank-based method used the implementation presented by Kuo {\it et al}.~\cite{kuo2014large}, and the SPDRank-based method used the implementation presented by Ohue {\it et al}.~\cite{ohue2019spdrank}.
The PKRank method searches for the cost parameter $C$ ($10^{-9}, 10^{-8}, ..., 10^0$) and the parameter $\gamma$ ($10^{-6}, 10^{-5}, 10^{-4}, 10^{-3}$) using an RBF kernel.

Figure~\ref{fig:concept}(a) shows the training and evaluation procedure for Dataset 1. The test data are those listed in Table~\ref{tab:exp1_test}. 
The validation data used for hyperparameter tuning does not contain information related to the proteins in the test. 
For example, when predicting the CHEMBL3706318 assay of PDE 2A as test data, 22 assays of PDE 7A and 2 assays of PDE 8B were used as validation data, and the assays of PDE 2A were not used.
The metric used for tuning parameters was NDCG@10.

Figure~\ref{fig:concept}(b) shows the procedure for Dataset 2.
For Dataset 2, data were randomly split into training and test sets in a $1:1$ ratio.
The training data size differs substantially from that of Dataset 1. Thus, the training data were sampled randomly so that the number of instances was 10000.
Moreover, the parameter tuning was performed by 5-fold cross-validation, which was sampled randomly so that the number of training instances for each fold was 10000.
The metric used for tuning parameters was NDCG$\%$10, representing the performance of the top 10\% of the prediction.

\section{Results}
\begin{figure*}[htbp]
    \centering
    \begin{subfigure}[htbp]{0.95\columnwidth}
        \centering
		\includegraphics[width=\columnwidth, bb= 0 0 550 400]{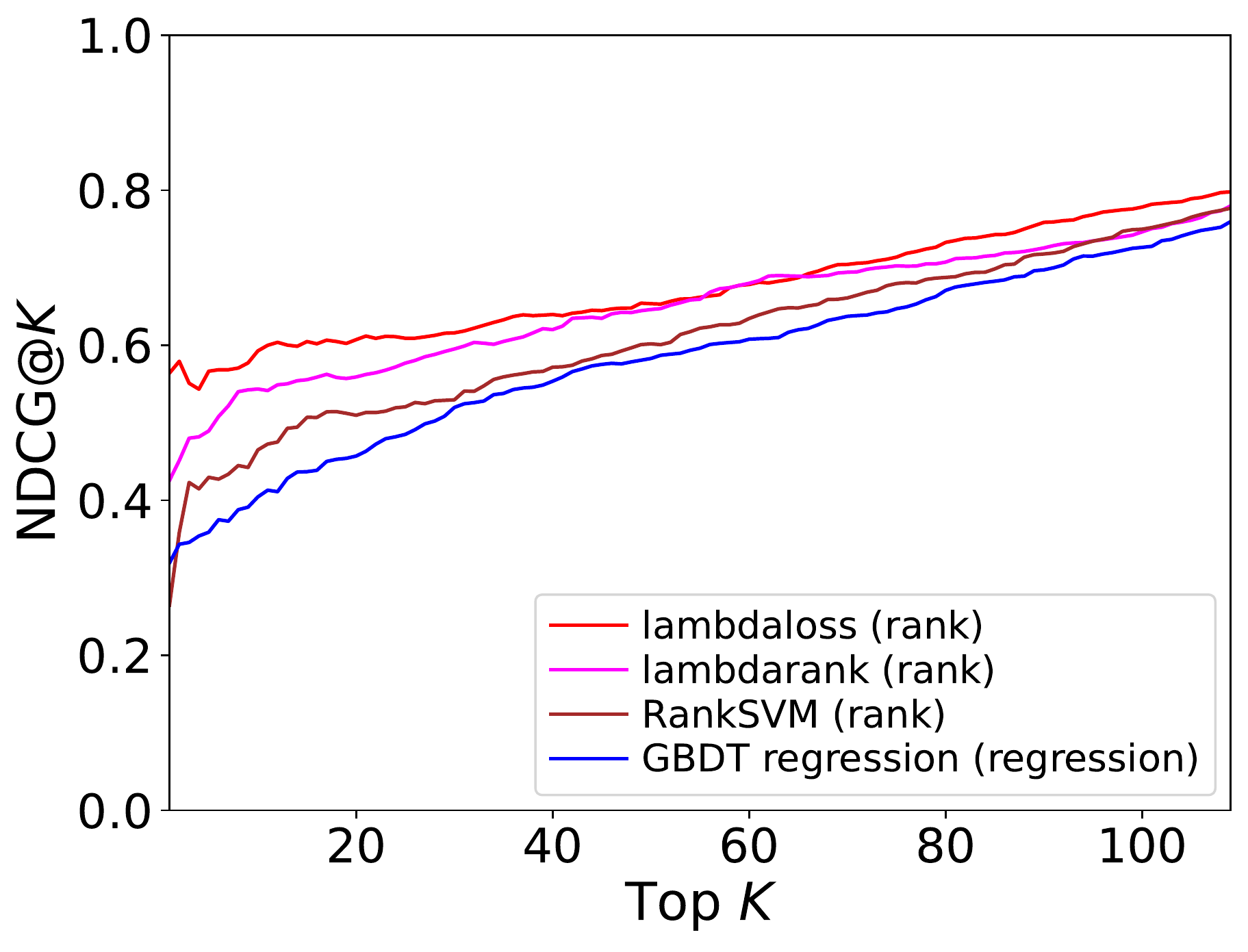}
        \subcaption{Dataset 1 (complicated task), NDCG}\vspace{6mm}
		\label{fig:exp1_results_NDCG}
    \end{subfigure}
    \hfill
    \begin{subfigure}[htbp]{0.95\columnwidth}
        \centering
		\includegraphics[width=\columnwidth, bb= 0 0 550 400]{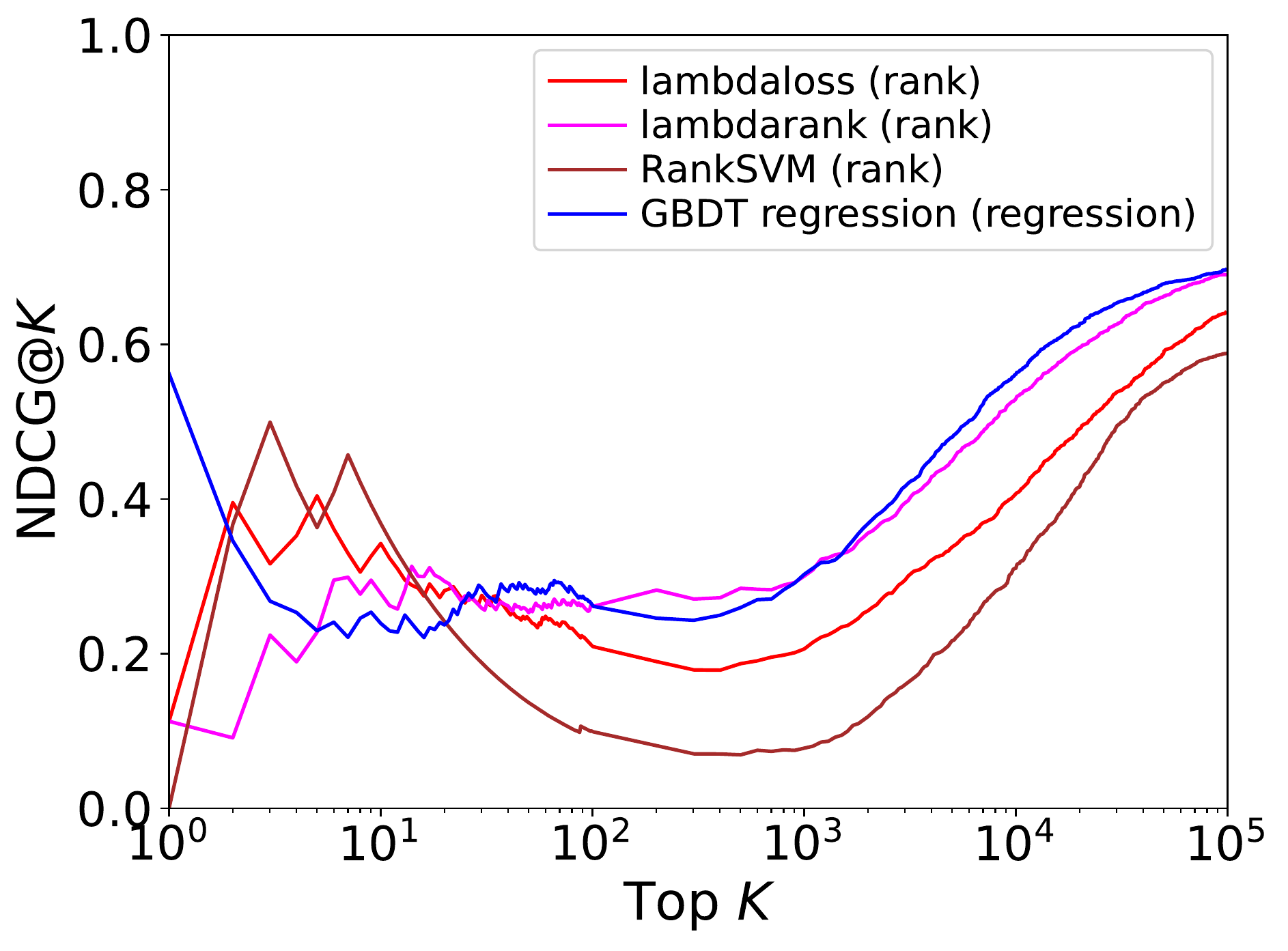}
        \subcaption{Dataset 2 (simple task), NDCG}\vspace{6mm}
		\label{fig:exp2_results_NDCG}
    \end{subfigure}
    \begin{subfigure}[htbp]{0.95\columnwidth}
        \centering
		\includegraphics[width=\columnwidth, bb= 0 0 550 400]{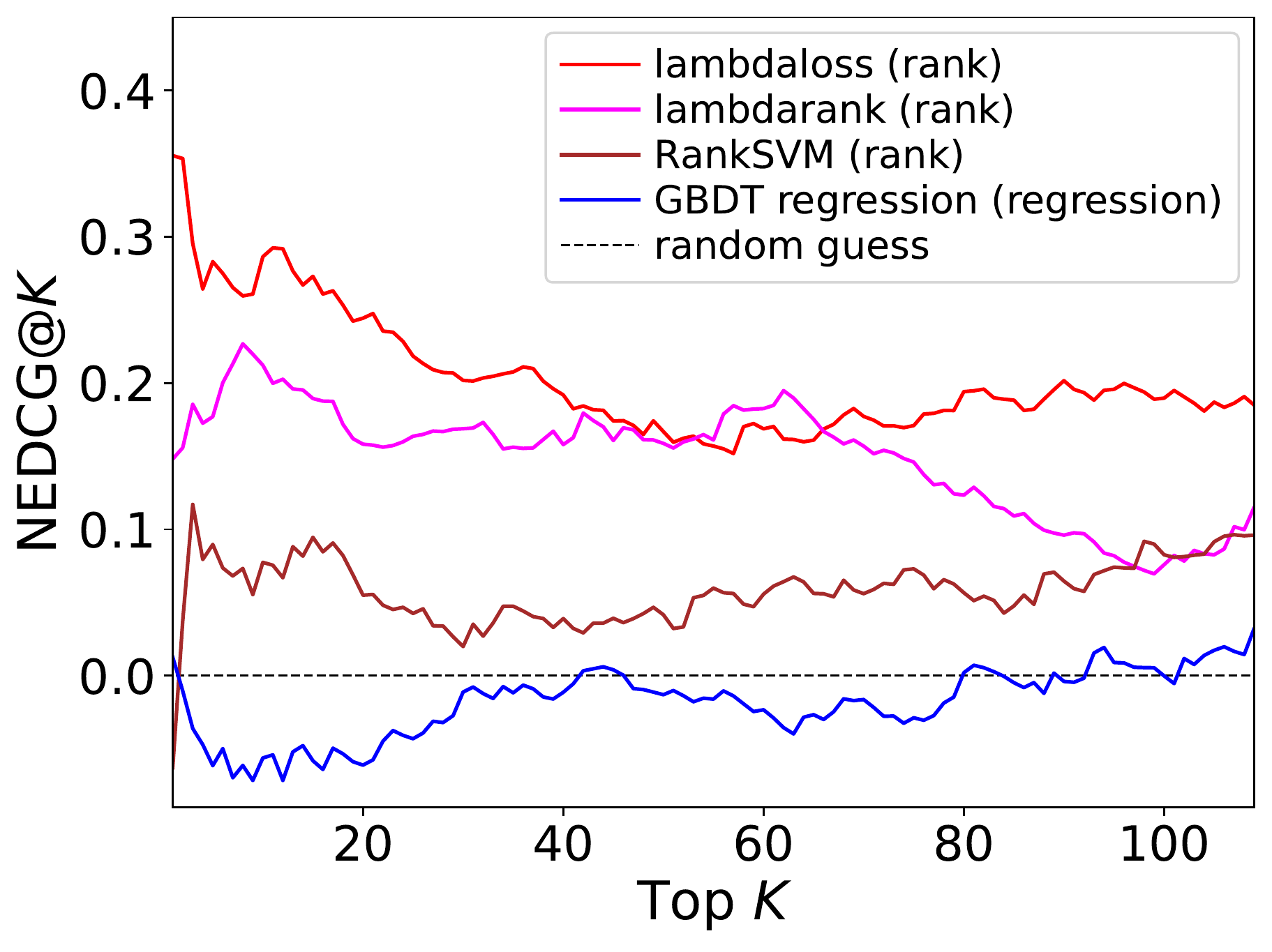}
        \subcaption{Dataset 1 (complicated task), NEDCG}\vspace{3mm}
		\label{fig:exp1_results_NEDCG}
    \end{subfigure}
    \hfill
    \begin{subfigure}[htbp]{0.95\columnwidth}
        \centering
		\includegraphics[width=\columnwidth, bb= 0 0 550 400]{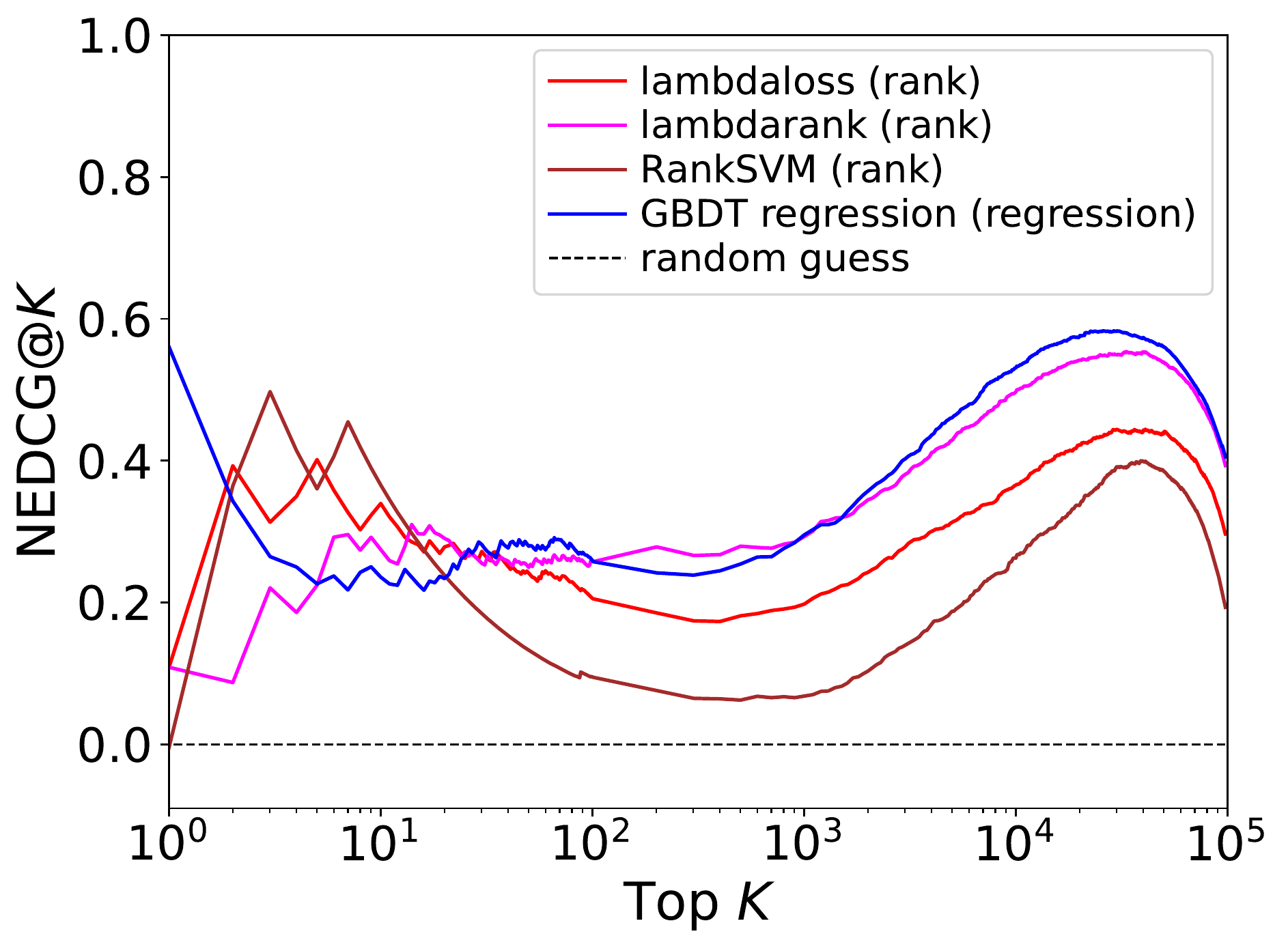}
        \subcaption{Dataset 2 (simple task), NEDCG}\vspace{3mm}
		\label{fig:exp2_results_NEDCG}
    \end{subfigure}
    \hfill
    \caption{Prediction results for Dataset 1 and Dataset 2. The evaluation metrics are NDCG and NEDCG of top $K$ samples, and the lines indicate the metrics for increasing values of $K$. The dashed lines in NEDCG plots represent random guessing (NEDCG = 0).}
    \label{fig:results}
\end{figure*}

\begin{figure*}[htbp]
    \centering
    \begin{subfigure}[htbp]{0.95\columnwidth}
        \centering
		\includegraphics[width=\columnwidth, bb= 0 0 600 380]{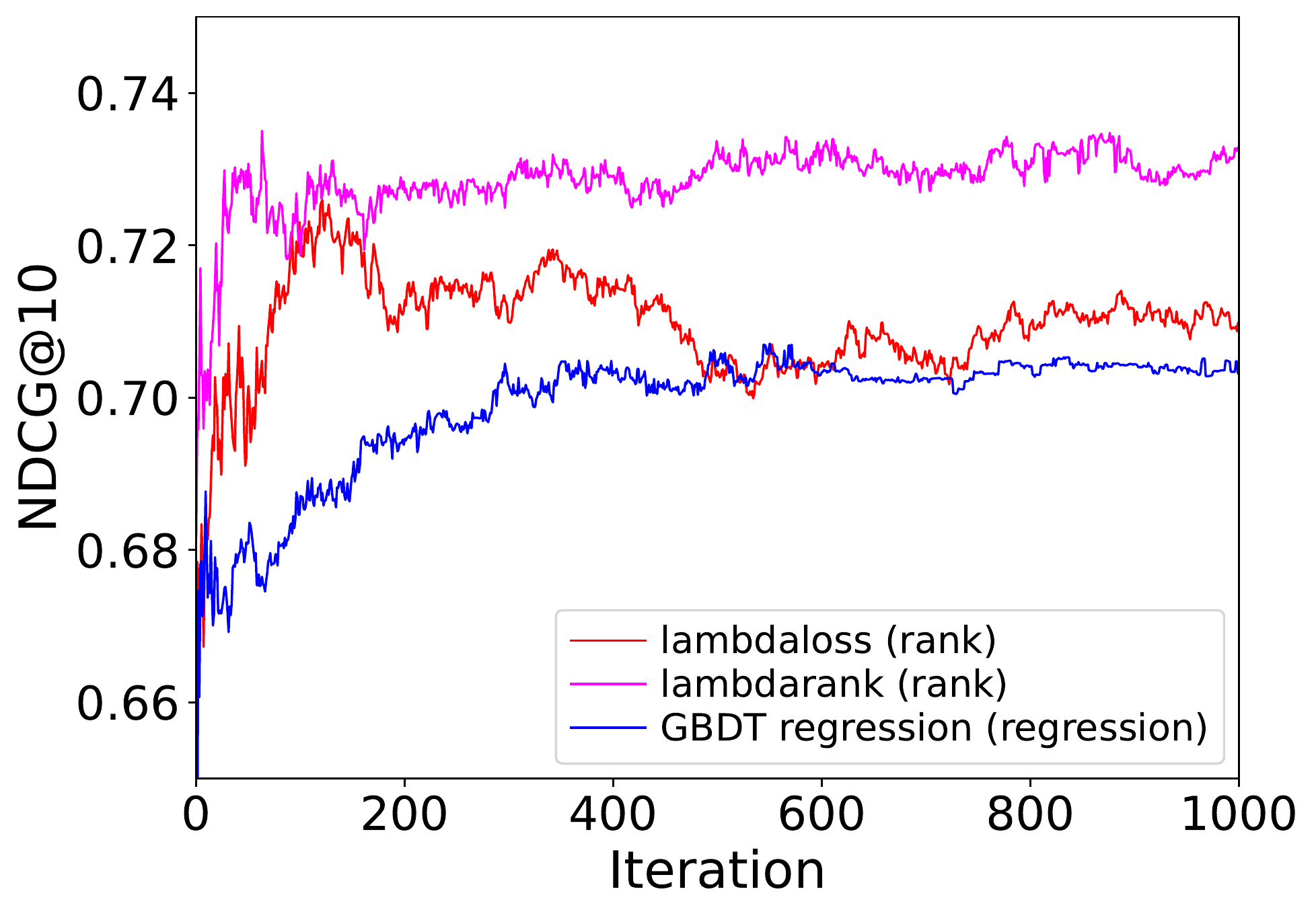}
        \subcaption[short for lof]{Dataset 1 (complicated task), NDCG@10}
		\label{fig:exp1_training_valid}
    \end{subfigure}
    \hfill
    \begin{subfigure}[htbp]{0.95\columnwidth}
        \centering
		\includegraphics[width=\columnwidth, bb= 0 0 600 380]{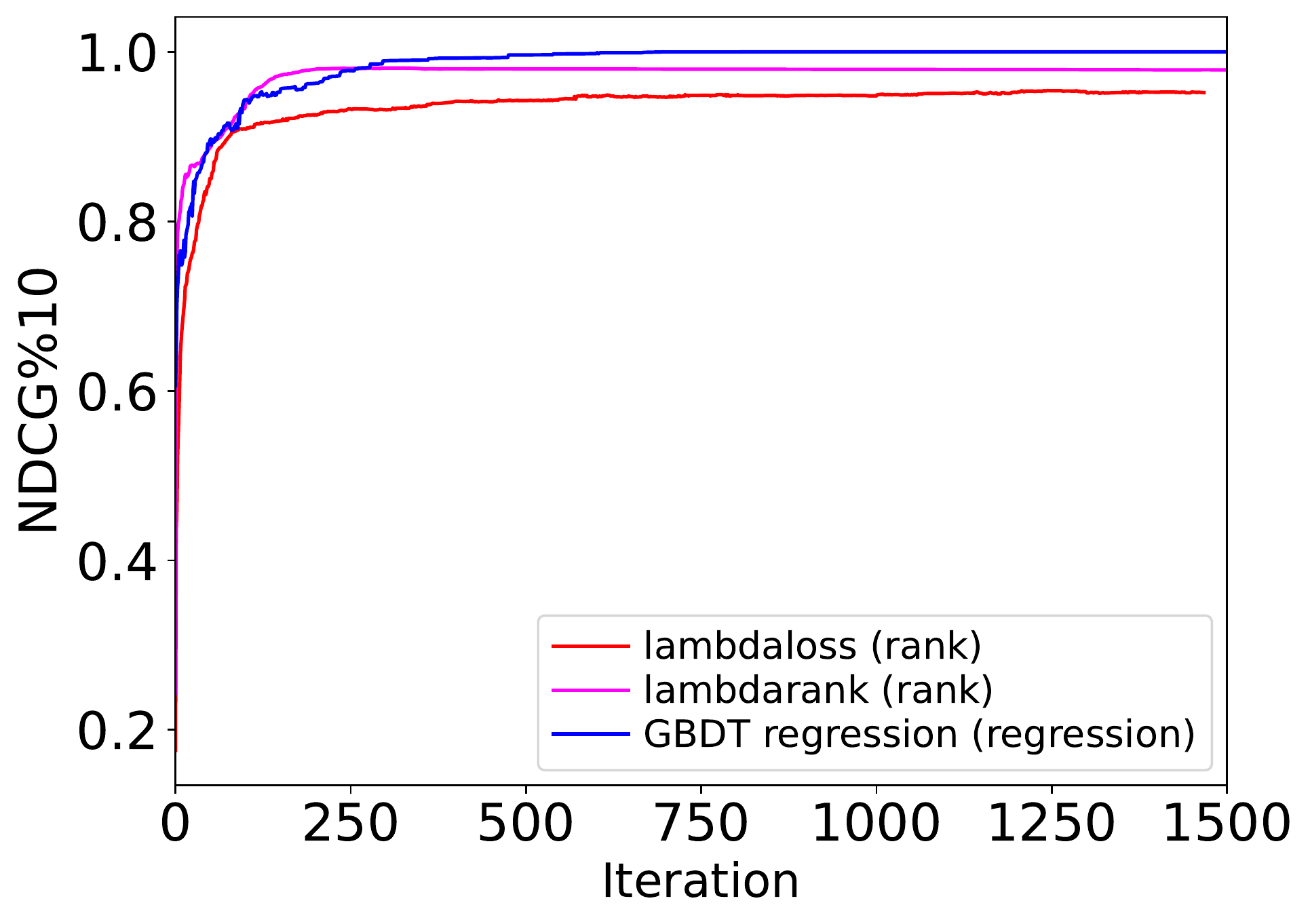}
        \subcaption[short for lof]{Dataset 2 (simple task), NDCG$\%$10}
		\label{fig:exp2_training_valid}
    \end{subfigure}
    \hfill
    \caption{NDCG for each iteration of the validation data during the training of the GBDT models. The figure shows (a) the values per iteration of NDCG@10 for Dataset 1 and (b) the values per iteration of NDCG$\%$10 for Dataset 2.}
    \label{fig:training}
\end{figure*}

\begin{table*}[tb]
    \centering
    \caption{ NDCG and NEDCG scores of the top 10 and top 10\%{} samples for Dataset 1 and Dataset 2 for each model. Boldface indicates the value for the method exhibiting the highest prediction accuracy for each dataset. Italics indicate methods judged to have negative values in NEDCG, i.e., worse than random guessing.}
    \begin{tabular}{l|rr|rr|rr|rr}
    \hline
    \hline
        ~ & \multicolumn{2}{c|}{lambdaloss (rank)} &
        \multicolumn{2}{c|}{lambdarank (rank)} &
        \multicolumn{2}{c|}{RankSVM (rank)} &
        \multicolumn{2}{c}{GBDT regression} \\ \cline{2-9}
        ~ & NDCG & NEDCG & NDCG & NEDCG & NDCG & NEDCG & NDCG & NEDCG \\ \hline
        Dataset 1 (top 10, $K=10$) & {\bf 0.593} & {\bf 0.286} & 0.543 & 0.212 & 0.465 & 0.077 & 0.404 & {\it $-$0.056} \\ 
        Dataset 1 (top 10\%, $K=11$) & {\bf 0.600} & {\bf 0.600} & 0.541 & 0.541 & 0.472 & 0.472 & 0.413 & 0.413 \\  \hline
        Dataset 2 (top 10, $K=10$) & 0.342 & 0.340 & 0.278 & 0.275 & {\bf 0.368} & {\bf 0.366} & 0.239 & 0.236 \\ 
        Dataset 2 (top 10\%, $K=9{,}778$) & 0.405 & 0.364 & 0.526 & 0.494 & 0.310 & 0.263 & {\bf 0.559} & {\bf 0.529} \\ \hline
    \end{tabular}
    \label{tab:result}
\end{table*}

Figure~\ref{fig:results} shows the experimental results for the 2 datasets using 2 metrics, NDCG@$K$ and NEDCG@$K$.
A higher score for a few samples indicates that the model can rank the active compounds at the top of the prediction.

Figure~\ref{fig:results}(a) shows that the NDCG@$K$ for the top dozen cases are lambdaloss, lambdarank, RankSVM, and regression, in that order.
This result indicates that learning-to-rank with GBDT provides more appropriate insights than existing RankSVM or GBDT with regression in a compound screening of novel targets using assays with various environments, such as Dataset 1.

In Figure~\ref{fig:results}(b), for the 97,782 test samples, the regression is best for NDCG@1, while RankSVM and lambdaloss are better for NDCG@10.
After that, no significant difference is present in the overall performance between lambdarank and regression.
These results indicate that learning-to-rank is less effective for single assays such as Dataset 2. Moreover, this method does not consistently outperform regression methods.
Note that RankSVM exhibited the lowest prediction accuracy; however, this was not an exact comparison, as RankSVM with fingerprinted features and linear kernels was used for Dataset 2.

In Figure~\ref{fig:results}(c), lambdaloss shows the best NEDCG value, followed by lambdarank, RankSVM, and regression, in that order.
Here, the NEDCG values for the regression method are generally equal to or less than 0.0, making it less accurate than random guessing.
Therefore, a learning-to-rank method rather than regression learning is appropriate for LBVS that uses data on different assay systems and multiple proteins.
As in Dataset 1, it is essential to consider how much the prediction of a new target has improved from a random prediction, and NEDCG is available for this assessment.

Figure~\ref{fig:results}(d) is almost the same as Figure~\ref{fig:results}(b).
This is because the prediction is improved at least a dozen times compared to the random prediction in each model.

Table~\ref{tab:result} summarizes the NDCG and NEDCG scores of the top 10 and top 10 $\%$ for Dataset 1 and Dataset 2 for each model, respectively.
The top 10 $\%$ comprises 11 samples for Dataset 1 and 9,778 samples for Dataset 2.
The scores for the top 10 samples indicate how well they predict compounds with good activity at the limited top of the prediction. The scores for the top 10 $\%$ indicate the performance of observing the top cases in the dataset.

\section{Discussion}
\subsection{Training of the GBDT}

Figure~\ref{fig:training} shows the convergence of validation scores in the training of the GBDT model.
The validation data in Dataset 1 consists of assay data for target proteins that the training data does not contain. Thus, the model overfits the training data as the number of iterations increases.
In contrast, the validation data in Dataset 2 are randomly split from the training data. 
We believe that the similar distribution of the training and validation data is why the scores did not decrease during training.
In this experiment, the early stopping parameter was set to 1000 rounds; thus, no concern regarding overfitting was present even with Dataset 1.

\subsection{NEDCG is an intuitive metric}
The proposed metric, NEDCG, is effective for complicated tasks such as Dataset 1.
In Dataset 1, the NEDCG of the GBDT regression is negative.
This score implies that GBDT regression predicts worse than random prediction.
If we had used NDCG, we would not have noticed this problem.
Using the NEDCG score, we could detect such hazards for the first time.
Thus, in a complicated task such as Dataset 1, examining whether the prediction accuracy is sufficient is crucial.
However, the difference between the NDCG and NEDCG values was slight because the Dataset 2 results improved substantially from the random predictions.
For such datasets, either NEDCG or NDCG provide similar results.
\subsection{Comparison of RankSVM and GBDT}
From Dataset 1, the GBDT ranking method was more accurate than RankSVM.
The RankSVM method learns ordinal relations for all pairs, whereas the methods using the lambdarank and lambdaloss loss functions aim to optimize NDCG directly.
In addition, the GBDT model is more practical than RankSVM for large datasets because the computation time of the RankSVM method increases in the order of the cube of the number of training instances.
Thus, the GBDT-based method with lambdarank and lambdaloss outperforms the RankSVM-based method regarding prediction accuracy for LBVS and their applicability to a large dataset.

\subsection{Regression vs. learning-to-rank}
\label{subsec:rank_vs_reg}
This section discusses when to use the learning-to-rank.
Although learning-to-rank is superior because it can integrate assay data from different environments to make predictions, as shown in the results section, this approach has the following disadvantages compared to regression.

RankSVM, lambdarank, and lambdaloss train in pairs or listwise, which can be slower than regression.
However, in learning-to-rank with GBDT, some techniques reduce the computational complexity, such as branch pruning with \texttt{lambdarank\_truncation\_level}, which is a parameter of LightGBM.
Similarly, SPDRank can reduce training time for large datasets by ignoring meaningless order~\cite{ohue2019spdrank}.
Therefore, no considerable difference in training time between regression and learning-to-rank methods was observed in most cases.

Thus, the ranking prediction should be used as an alternative when regression prediction cannot achieve practical accuracy, for example, in Dataset 1.
The information on activity values obtained in regression is lost in the ranking score.
The experimental results with a single assay in Dataset 2 showed little effect of learning-to-rank, and the predictions from regression learning seem to be somewhat reliable.
Conversely, in Dataset 1, the regression method was comparable to random prediction, indicating that learning-to-rank was effective for datasets with multiple assays and targets.
If the NEDCG is near 0.0, we can quantitatively judge that the model is not predicting adequately. Therefore, we can answer when to use learning-to-rank based on the newly proposed evaluation metric, NEDCG.

\section{Conclusion}
We evaluated the ranking prediction performance of the GBDT model using a lambdarank loss function aimed to directly optimize the NDCG to improve the accuracy of VS through learning-to-rank.
The comparison between the proposed method with RankSVM and GBDT regression models showed that the proposed method outperformed the other models on a dataset that integrated biochemical assay data from various environments, a case generally considered challenging for regression prediction.
We also examined the range of effectiveness of learning-to-rank and found that the effectiveness was not superior to regression methods for all ranking predictions and that learning-to-rank was particularly effective in situations where data integration was essential.
In addition, the proposed NEDCG provided a more intuitive evaluation of prediction favorability than the existing NDCG because a value of NEDCG greater than or equal to 0 indicates that the prediction is better than a random prediction.

We focused on the GBDT method using learning-to-rank. However, deep learning models, which have been gaining popularity in recent years, may further improve the performance of LBVS using learning-to-rank~\cite{jiang2021could}.
Although learning-to-rank with deep learning models is not necessarily superior to descriptor-based machine learning methods such as GBDT and RankSVM, an examination of their limitations is required.

Furthermore, learning-to-rank may apply to other tasks in chemoinformatics besides affinity prediction,
such as ADMET (absorption, distribution, metabolism, excretion, and toxicity) prediction~\cite{10.1093/bioinformatics/btab726}, QSAR (Quantitative Structure-Activity Relationship)~\cite{matsumoto2021ranking}, and drug-target interaction prediction~\cite{10.1093/bioinformatics/btac048}.
For these tasks, ranking predictions may be practical for small datasets for regression prediction or when integrating assay data from various environments.

\section*{Acknowledgment}
The authors thank Yutaka Akiyama and Keisuke Yanagisawa at the Tokyo Institute of Technology for their constructive discussion and feedback. This work was financially supported by Japan Science and Technology Agency (JST) FOREST (Grant No. JPMJFR216J), JST ACT-X (Grant No. JPMJAX20A3), Japan Society for the Promotion of Science (JSPS) KAKENHI (Grant No. 20H04280), and Japan Agency for Medical Research and Development (AMED) Basis for Supporting Innovative Drug Discovery and Life Science Research (BINDS) (Grant No. JP22ama121026).

\bibliographystyle{IEEEtran}
\bibliography{mainbib}
\end{document}